\documentclass[twoside,a4paper]{IEEEtran}

\usepackage[IEEEtran]{research18}

\begin{document}

\date{}

\newcommand{\mH}{\mathcal{H}}
\newcommand{\Db}{D_\textnormal{b}}
\renewcommand{\Hb}{H_\textnormal{b}}
\newcommand{\mE}{\mathcal{E}}

\title{Improved Random-Binning Exponent for \\Distributed Hypothesis Testing}

\author{
    Yuval Kochman and Ligong Wang
    \thanks{Yuval Kochman is with the School of Computer Science and Engineering, the Hebrew University of Jerusalem, Israel (email: yuvalko@cs.huji.ac.il). Ligong Wang is with the Department of Information Technology and Electrical Engineering, ETH Zurich, Switzerland (e-mail: ligwang@isi.ee.ethz.ch).}}

\maketitle

\begin{abstract}
  Consider the problem of distributed binary hypothesis testing with two terminals, where the decision is made at one of them (the ``receiver''). We study the exponent of the error probability of the second type. Previously, an achievable exponent was derived by Shimokawa, Han, and Amari using a ``quantization and binning'' scheme. We propose a simple modification on the receiver’s decision rule in this scheme to attain a better exponent. 
\end{abstract}


\section{Introduction}
In distributed hypothesis testing, one wishes to distinguish between different possible joint distributions of data observed at several terminals, when communication between the terminals is rate-limited. It is a classic problem that lies in the intersection of statistics and information theory, and that continues to attract attention from researchers; see, e.g., \cite{ahlswedecsiszar86,han87,shalabypapamarcou92,shimokawahanamari94,rahmanwagner12,zhaolai18,weinbergerkochman19,salehwiggerwang19,salehwigger20,sreekumargunduz20,watanabe22}. 

Here, we study the simple setting with two observers who observe outcomes of random sequences $X^n$ and $Y^n$, respectively. We consider the discrete memoryless case, so both alphabets $\set{X}$ and $\set{Y}$ are finite, and $(X^n,Y^n)$ are independent and identically distributed (IID) over time. Under the null hypothesis $\mH_0$ their joint probability mass function at any time instant is $P_{XY}$, and under the alternative hypothesis $\mH_1$ it is $Q_{XY}$:
\begin{IEEEeqnarray}{rCl}
\mH_0\colon &\quad& (X^n,Y^n)\sim \textnormal{IID }P_{XY}\\
\mH_1\colon &\quad& (X^n,Y^n)\sim \textnormal{IID }Q_{XY}.
\end{IEEEeqnarray}
We shall focus on the scenario where the decision is made by one of the observers. Specifically, we refer to the observer who observes $X^n$ as the \emph{sender} and the one who observes $Y^n$ as the \emph{receiver}. Let $R>0$ denote the permitted communication rate. The sender creates an $n R$-bit message $m=f(x^n)$ and sends it to the receiver. Then the receiver makes a decision between $\mH_0$ and $\mH_1$ using some function $g(m,y^n)$. 

We require the probability of a decision error by the receiver under $\mH_0$ to approach zero (possibly slowly), and seek the fastest decay of error probability under $\mH_1$. More formally, let $p(\epsilon,n,R)$ denote the smallest attainable error probability under $\mH_1$ when the observation length is $n$, the permitted communication rate is $R$, and the error probability under $\mH_0$ is required to be less than or equal to $\epsilon$. We are interested in
\begin{equation}
E(R) \triangleq \lim_{\epsilon\downarrow 0} \lim_{n\to\infty} -\frac{1}{n} \log p(\epsilon,n,R).
\end{equation}

The exact characterization of $E(R)$ for general $P_{XY}$ and $Q_{XY}$ is a long-standing open problem. Well known are three schemes and corresponding lower bounds on $E(R)$, respectively, by Ahlswede and Csisz\'ar (AC) \cite{ahlswedecsiszar86}, by Han \cite{han87}, and by Shimokawa, Han, and Amari (SHA) \cite{shimokawahanamari94}. The idea of both AC and Han is for the sender to produce a lossy compression $u^n$ of its observation $x^n$, and to send the index of $u^n$ to the receiver; Han's analysis yields a better error exponent than AC. 
SHA use random binning as in Wyner-Ziv coding \cite{wynerziv76}, which allows the sender to compress $x^n$ at rates that are larger than $R$, and to send only part of the index of $u^n$ to the receiver. 

AC's error exponent is optimal for ``testing against independence,'' namely, when 
$Q_{XY}= P_X \cdot P_Y$ \cite{ahlswedecsiszar86}.  SHA's exponent is optimal for a scenario called ``testing against conditional independence'' (which includes ``testing against independence'' as a special case); see Rahman and Wagner \cite{rahmanwagner12}. 

Weinberger and Kochman \cite{weinbergerkochman19} consider the encoding scheme of SHA to analyze Neyman-Pearson tests \cite{coverthomas06} by the receiver that attain exponential decay in the error probabilities under both $\mH_0$ and $\mH_1$. Their analysis can be specified to a single error exponent by setting the other error exponent to zero. Although it is conceivable that the resulting single error exponent may be better than that of SHA (due to the optimality of Neyman-Pearson tests), the expression appears difficult to evaluate.

Watanabe \cite{watanabe22} constructs an example where all the above schemes are suboptimal: a better error exponent can be obtained when one applies two SHA-type schemes in parallel.

In this work, we propose a simple and natural improvement on SHA's scheme---specifically, its decision rule. 
The new error exponent is at least as good as SHA's exponent, while in some cases it is strictly larger. We shall present the new scheme in Section~\ref{sec:main}, after first reviewing Han's and SHA's schemes in the next section.

\subsection*{Some Notation}

Throughout this paper, by ``$\epsilon$-typical,'' $\epsilon>0$, we refer to the definition given in \cite[Section~2.4]{elgamalkim11} (some literature calls it ``robust typicality''). We may omit the parameter $\epsilon$ when it is clear from the context. 

In all schemes below, we fix a finite auxiliary set $\set{U}$ and a stochastic kernel $P_{U|X}$. Denote 
\begin{subequations}\label{eq:defPUXY}
\begin{IEEEeqnarray}{rCl}
P_{UXY} & = &P_{U|X} P_{XY} \\
 Q_{UXY} & = & P_{U|X} Q_{XY}.
\end{IEEEeqnarray}
\end{subequations}
We sometimes add subscripts to mutual informations as in $I_Q(U;Y)$ to specify the distributions under which they are computed. When there are no subscripts, it shall be understood that they are computed under $P_{UXY}$.

\section{A Brief Review}

\subsection{Han's Scheme}

Generate a codebook of sequences
\begin{equation}
u^n(m),\qquad m\in\{1,\ldots,2^{nR}\}
\end{equation}
each IID according to $P_U$---the $U$-marginal of $P_{UXY}$---and independently of each other. 

\medskip

\emph{Sender (Han):} Upon observing $x^n$, look for an index $m$ such that $(u^n(m),x^n)$ are jointly $\epsilon$-typical, $\epsilon>0$, according to $P_{UX}$. If such indices can be found, send any one of them to the receiver; if no such index can be found, send a special message to the receiver indicating ``Declare~$\mH_1$.'' 

\medskip

\emph{Receiver (Han):} If the special message is received, declare $\mH_1$. If an index $m$ is received, check whether or not $(u^n(m),y^n)$ are jointly $\epsilon'$-typical according to $P_{UY}$, where $\epsilon'>\epsilon$. If they are, then declare $\mH_0$; otherwise declare $\mH_1$. 

\medskip

Under $\mH_0$, to ensure that the sender can find a good codeword $u^n(m)$ with high probability, we require 
\begin{equation} \label{eq:RIUX}
R > I(U;X),
\end{equation}
where, as we recall, $I(U;X)$ is computed according to $P_{UX}$.
Provided that $(u^n(m),x^n)$ are jointly typical, the probability of a decision error by the receiver is guaranteed to tend to zero as $n\to\infty$ by the Conditional Typicality Lemma \cite[Section~2.5]{elgamalkim11}. 

Under $\mH_1$, an error occurs if there exists $u^n(m)$ that is jointly typical with $x^n$ according to $P_{UX}$ (which requires $x^n$ itself to be typical according to $P_X$), and $(u^n(m),y^n)$ happen to be jointly typical according to $P_{UY}$. The exponent of this probability (when $\epsilon$ and $\epsilon'$ are both made to approach zero) can be computed to be
\begin{equation}\label{eq:EHan}
E_0(P_{U|X}) \triangleq \min_{\hat{P}_{UXY}\in\set{P}_{\textnormal{Han}(P_{UXY})}} D\left( \hat{P}_{UXY} \middle\| Q_{UXY} \right)
\end{equation}
with
\begin{equation}
\set{P}_{\textnormal{Han}} (P_{UXY})\triangleq \left\{ \hat{P}_{UXY}\colon \hat{P}_{UX} = P_{UX},\hat{P}_{UY}=P_{UY}\right\}.
\end{equation}
Hence Han's scheme can achieve any error exponent below
\begin{equation}
E_\textnormal{Han} (R)= \sup_{\substack{P_{U|X}\colon \\ I(U;X)< R}} E_0(P_{U|X}).
\end{equation}

\subsection{Binning and SHA's Scheme}\label{sec:binning}

Viewing $y^n$ as side information for the receiver, the sender can use binning as in Wyner-Ziv coding \cite{wynerziv76}. 
Fix some $R'>0$ and generate a codebook IID according to $P_U$
\begin{equation}\label{eq:bin_codebook}
u^n(m,\ell),\qquad m\in\{1,\ldots,2^{nR} \}, \quad \ell \in\{1,\ldots,2^{nR'}\}.
\end{equation}

\medskip

\emph{Sender (binning):} Upon observing $x^n$, look for a pair $(m,\ell)\in\{1,\ldots,2^{nR}\}\times\{1,\ldots,2^{nR'}\}$ such that
\begin{equation}
\left( u^n(m,\ell), x^n \right) \textnormal{ are jointly $\epsilon$-typical according to $P_{UX}$.}
\end{equation}
If successful, send any such $m$. If unsuccessful, send a special message ``Declare $\mH_1$.'' 

\medskip

The requirement on the size of the codebook is the same as in Han's scheme, except the size is now given by $2^{n(R+R')}$ instead of $2^{nR}$. Thus we require
\begin{IEEEeqnarray}{rCl} \label{eq:RR'}
R+R' & > & I(U;X).
\end{IEEEeqnarray}

In the rest of this section we shall focus on cases where $P_{XY}$ and $Q_{XY}$ have the same marginals, i.e.,
\begin{subequations}\label{eq:samemarginal}
\begin{IEEEeqnarray}{rCl}
P_X  & = & Q_X,\\
 P_Y & = &Q_Y.
\end{IEEEeqnarray}
\end{subequations}
This assumption simplifies the discussion while still capturing the essence of the difference between the schemes. 
When \eqref{eq:samemarginal} does not hold, SHA's error exponent remains valid in its form that we later give. But, before discussing SHA's receiver, we first introduce a ``na\"ive'' receiver. We note that the error exponent of the na\"ive receiver will need to be modified if we do not assume \eqref{eq:samemarginal}.


\medskip

\emph{Receiver (na\"ive):} Upon receiving an index $m$, look for $\hat{\ell}$ such that $(u^n(m,\hat{\ell}),y^n)$ are jointly $\epsilon'$-typical according to $P_{UY}$, $\epsilon'>\epsilon$. If such an $\hat{\ell}$ can be found, declare $\mH_0$. Otherwise declare $\mH_1$. (If the special message is received, also declare~$\mH_1$.)

\medskip

As long as $\left(u^n(m,\ell), x^n \right)$ are jointly typical (which happens with high probability as long as \eqref{eq:RR'} holds), the error probability under $\mH_0$ is guaranteed to tend to zero as $n\to\infty$, again by the Conditional Typicality Lemma.

Under $\mH_1$, there are two types of errors. The first type is $(u^n(m,\ell), y^n)$ are jointly typical according to $P_{UY}$. The exponent of this error probability is $E_0(P_{U|X})$ as in Han's scheme. The second type is, for some $k\neq \ell$, $(u^n(m,k),y^n)$ are jointly typical according to $P_{UY}$. For every $k$, since $U^n(m,k)$ and $Y^n$ are generated independently, the probability that they are jointly typical according to $P_{UY}$ is approximately $2^{- n I(U;Y)}$. Since there are $(2^{nR'}-1)$ possibilities for $k$, the probability of the second error type is approximately
\begin{equation} \label{eq:R'}
2^{-n ( I(U;Y) - R')^+}.
\end{equation}
Summarizing the above and recalling that, due to \eqref{eq:RR'}, $R'$ can take values up to $I(U;X)-R$, we conclude that the error exponent of this scheme is given by
\begin{IEEEeqnarray}{rCl}
\IEEEeqnarraymulticol{3}{l}{
E_{\textnormal{na\"ive}} (R) }\nonumber\\*
~~& = & \sup_{\substack{P_{U|X}\colon \\ I(U;X|Y)< R < I(U;X)}} \min \left\{ E_0 (P_{U|X}),\,\, R-I(U;X|Y) \right\}.\nonumber\\*
\label{eq:naive}
\end{IEEEeqnarray}


\medskip

With \emph{Sender (binning)} above, let us now suppose that the receiver finds two indices $\ell_1, \ell_2$ such that $(u^n(m,\ell_1),y^n)$ are typical according to $P_{UY}$ and $(u^n(m,\ell_2),y^n)$ are typical according to $Q_{UY}$. Should it declare $\mH_0$ or $\mH_1$? Our na\"ive receiver would always declare $\mH_0$, but a more clever receiver should consider the following question: What is better to assume, that $u^n(m,\ell_1)$ is the correct codeword (i.e., $\ell_1=\ell$) and $u^n(m,\ell_2)$ is generated independently of $y^n$, or the other way around? If $I_P(U;Y)< I_Q(U;Y)$, then an ``incorrect'' $u^n$-codeword and $y^n$ being jointly $P_{UY}$-typical is more likely than their being $Q_{UY}$-typical, suggesting that the receiver should declare $\mH_1$ in such a scenario. 

SHA's receiver takes the above observation into account by first decoding the codeword chosen by the sender in a ``universal'' manner. Specifically, since the receiver does not know the actual joint distribution (which could be either $P$ or $Q$), it picks the codeword whose joint empirical distribution with $y^n$ minimizes $H(U|Y)$ among all codewords in the bin.\footnote{This is related to the \emph{maximum mutual information decoder} used in universal channel coding; see \cite[Chapter 10]{csiszarkorner11}.} It then checks whether or not this codeword and $y^n$ are $P_{UY}$-typical. 

Here we describe SHA's receiver in a slightly different way, which is equivalent to the original one.

\medskip

\emph{Receiver (SHA):} Declare $\mH_0$ if an index $m$ is received, there exists $\hat{\ell}\in\{1,\ldots,2^{nR'}\}$ such that $(u^n(m,\hat{\ell}),y^n)$ are jointly typical according to $P_{UY}$, and the following is true:
\begin{equation} \label{eq:SHAcondition}
	I_{\pi^k}(U;Y) \le I_P(U;Y)\quad \textnormal{for all }k\neq \hat{\ell},
\end{equation}
where $\pi^{k}$ denotes the joint type \cite{csiszarkorner11} of $(u^n(m,k),y^n)$. Otherwise, declare $\mH_1$.

\medskip

Since \cite{shimokawahanamari94} does not contain a proof, we refer the reader to \cite[Appendix B]{salehwiggerwang19} for a detailed derivation of SHA's error exponent, which is given by
\begin{IEEEeqnarray}{rCl}
\IEEEeqnarraymulticol{3}{l}{
E_\textnormal{SHA}(R)}\nonumber\\*
~~~~&  =  &\sup_{\substack{P_{U|X}\colon \\ I(U;X|Y)<R< I(U;X)}} \min \Big\{ E_0(P_{U|X}), \,\, E_1(P_{U|X}, R)\Big\}, \nonumber \\* 
\end{IEEEeqnarray}
where the second term in the minimization is
\begin{IEEEeqnarray}{rCl}
\IEEEeqnarraymulticol{3}{l}{
E_1(P_{U|X}, R)}\nonumber\\*
~~~~~& \triangleq & \min_{\tilde{P}\in\set{P}_\textnormal{SHA}(P_{UXY})} D \left( \tilde{P}_{UXY}\middle\| Q_{UXY}\right) + R - I(U;X|Y) \nonumber\\*  \label{eq:SHAE1}
\end{IEEEeqnarray}
with 
\begin{IEEEeqnarray}{rCl} 
\set{P}_\textnormal{SHA}(P_{UXY}) & \triangleq & \Big\{ \tilde{P}_{UXY}\colon \tilde{P}_{UX}=P_{UX},\, \tilde{P}_{Y}=P_Y,\, \nonumber\\*
&& \qquad\qquad\quad I_{\tilde{P}}(U;Y)\le I_P(U;Y) \Big\}. \IEEEeqnarraynumspace
\label{eq:PSHA}
\end{IEEEeqnarray}

\medskip

It should be noted that, when $R$ approaches $I(U;X)$, $E_1(P_{U|X}, R)$ does \emph{not} necessarily approach or exceed $E_0(P_{U|X})$. Consequently, $E_\textnormal{SHA} (R)$ (as we define it) can be smaller than $E_\textnormal{Han}(R)$.\footnote{The exponents in SHA \cite{shimokawahanamari94} are defined slightly differently from ours. The eventual exponent in \cite{shimokawahanamari94} is, in our notation, $\max\{E_\textnormal{Han}(R),E_\textnormal{SHA}(R)\}$. } 

\section{The New Scheme}\label{sec:main}

\subsection{Some Intuition}\label{sub:intuition}
In (our interpretation of) SHA's receiver, every $\pi^k$ is compared with $P_{UY}$ in terms of mutual information. Let us consider using other functions for comparison. For example, we could impose a condition (for every $k\neq \hat{\ell}$) in terms of total variation distance: 
\begin{equation}\label{eq:TV}
 \delta_\textnormal{TV}(\pi^k,P_U \cdot P_Y) \le \delta_\textnormal{TV}(P_{UV},P_U \cdot P_Y)
\end{equation}
or relative entropy conditional on a specific $u\in\set{U}$:
\begin{equation}\label{eq:Du}
D(\pi_{Y|U=u}^k \| P_Y) \le D(P_{Y|U=u}\| P_Y).
\end{equation}
Each such condition will result in the constraint $I_{\tilde{P}}(U;Y)\le I_P(U;Y)$ in \eqref{eq:PSHA} being replaced by another constraint that corresponds to the condition that we choose. We could even impose several such conditions (including \eqref{eq:SHAcondition} itself) at the same time, resulting in a smaller set than $\set{P}_\textnormal{SHA}(P_{UXY})$ and hence possibly a larger error exponent under $\mH_1$.

However, changing \eqref{eq:SHAcondition} or adding more conditions may add restrictions on the bin size, i.e., on $R'$. Indeed, the probability for some codeword $u^n(m,k)$, $k\neq \ell$, to violate the imposed conditions must tend to zero as $n$ grows to infinity, otherwise the decision error probability under $\mH_0$ cannot tend to zero.\footnote{With SHA's condition \eqref{eq:SHAcondition}, this means $R' < I(U;Y)$, which is the same as the condition for the exponent in \eqref{eq:R'} to be nontrivial. Hence effectively \eqref{eq:SHAcondition} does not add any restriction on $R'$.} The right question to ask is therefore the following:
\medskip
\begin{quote}
\emph{Given $R'$, what are the strictest conditions that one can impose in place of~\eqref{eq:SHAcondition}, such that the error probability under $\mH_0$ will still tend to zero?}
\end{quote}
\medskip
First observe that the conditions we seek should only depend on the joint type of $(u^n(m,k),y^n)$, because the joint type determines the probability of the pair---be it computed under $P_{UY}$, $Q_{UY}$, or $P_U\cdot P_Y$. Further note that the probability for $(U^n,Y^n)$ to be of type $\pi$ under $P_U\cdot P_Y$ is approximately $2^{-nI_{\pi}(U;Y)}$ \cite{csiszarkorner11}. Since there are $2^{nR'}-1$ ``incorrect'' codewords (i.e., not the one chosen by the sender) in the bin, it follows that every type $\pi$ with $I_{\pi}(U;Y)> R'$ is highly unlikely to result from any ``incorrect'' codeword, and hence can and should be excluded by the conditions that we seek. Conversely, we cannot exclude those types with $I_{\pi}(U;Y)< R'$, because the probability for such types to ``randomly occur'' in a bin does not vanish, so excluding them will cause large error probability under~$\mH_0$. In other words, the condition that we seek is $I_{\pi^k}(U;Y) < R'$.

\subsection{Scheme and Result}
For the new scheme, we drop the same-marginal assumption \eqref{eq:samemarginal} to consider general $P_{XY}$ and $Q_{XY}$. As before, we fix an auxiliary set $\set{U}$ and a stochastic kernel $P_{U|X}$, and define $P_{UXY}$ and $Q_{UXY}$ as in \eqref{eq:defPUXY}. We require that, under $P_{UXY}$,
\begin{equation}\label{eq:IUXYR}
R < I(U;X).
\end{equation} 
Fix some $R'>0$. The codebook \eqref{eq:bin_codebook} is generated IID according to $P_U$, and the sender is \emph{Sender (binning)} from Section~\ref{sec:binning}.

\medskip

\emph{Receiver (new):} Declare $\mH_0$ if an index $m$ is received, and if both of the following are true:
\begin{enumerate}
\item For some $\epsilon'>\epsilon$, there exists $\hat{\ell}\in\{1,\ldots,2^{nR'}\}$ such that
\begin{equation}\label{eq:uyjt}
\big(u^n(m,\hat{\ell}),y^n\big) \textnormal{ are jointly $\epsilon'$-typical according to }P_{UY};
\end{equation}
\item For all $k \neq \hat{\ell}$, 
\begin{equation} \label{eq:Icond}
I_{\pi^k}(U;Y) < R'+\delta,
\end{equation} 
where $\delta>0$ will be chosen to approach zero later on, and where $\pi^k$ denotes the joint type of $(u^n(m,k),y^n)$.
\end{enumerate}
In all other cases, declare $\mH_1$.

\medskip

\begin{theorem}
The new scheme can achieve any error exponent that is below
\begin{equation}
\sup_{\substack{P_{U|X}\colon R < I(U;X)}} \min \left\{ E_0(P_{U|X}),\,\, E^*(P_{U|X},R)\right\},
\end{equation}
where $E_0(P_{U|X})$ is given in \eqref{eq:EHan} and 
\begin{IEEEeqnarray}{rCl}
E^*(P_{U|X},R) & \triangleq & \min_{\tilde{P}\in\set{P}^*(P_{UXY},\, I(U;X)-R)} D\left(\tilde{P}_{UXY}\middle\| Q_{UXY}\right) \nonumber\\
& & \qquad\qquad {} + \big(R - I(U;X|Y)\big)^+
\label{eq:EKW}
\end{IEEEeqnarray}
with
\begin{IEEEeqnarray}{rCl}
\set{P}^*(P_{UXY}, \tilde{R}) & \triangleq & \Big\{ \tilde{P}_{UXY}\colon \tilde{P}_{UX}=P_{UX},\, \tilde{P}_Y=P_Y,\nonumber\\*
& & \qquad\qquad \qquad\qquad I_{\tilde{P}}(U;Y) \le \tilde{R} \Big\}. \IEEEeqnarraynumspace
\label{eq:defP*}
\end{IEEEeqnarray}
\end{theorem}

\medskip

\begin{IEEEproof} We analyze different types of error that may occur. 


\emph{Error under $\mH_0$.} To ensure high probability for the sender to succeed in finding a codeword that is jointly typical with $x^n$, we require
\eqref{eq:RR'} to hold.  Assuming encoding is successful, there are two types of decision errors under $\mH_0$. The first is where the correct codeword $u^n(m,\ell)$ is not jointly typical with $y^n$ according to $P_{UY}$, the probability of which is guaranteed to be small by the Conditional Typicality Lemma \cite[Section~2.5]{elgamalkim11}. 
The second type of decision error is where some $u^n(m,k)$, $k\neq \ell$, and $y^n$ have empirical mutual information that is larger than $R'+\delta$. Note that $U^n(m,k)$ and $Y^n$ are drawn IID according to $P_U\cdot P_Y$. By Sanov's Theorem \cite[Theorem~11.4.1]{coverthomas06}, for every $k\neq{\ell}$, the probability for this to happen is upper-bounded by
\begin{equation}
(n+1)^{|\set{U}||\set{Y}|} 2^{-n D^*}
\end{equation}
where 
\begin{IEEEeqnarray}{rCl}
D^* & = & \min_{P_{UY}' \colon I_{P'}(U;Y) \ge R'+\delta} D\left(P_{UY}'  \middle\| P_U\cdot P_Y\right) \\
& = & \min_{P_{UY}' \colon I_{P'}(U;Y) \ge R'+\delta} I_{P'}(U;Y) + D\left(P_{U}'\cdot P_Y'  \middle\| P_U\cdot P_Y\right)\nonumber\\* \\
& = & R'+\delta.
\end{IEEEeqnarray}
Since the bin size is only $2^{nR'}$, by the union bound, the probability of this second error type is guaranteed to tend to zero as $n\to\infty$ for all $\delta>0$.

\emph{Error under $\mH_1$.}  
There are again two types of errors. The first is where the sender could find a codeword $u^n(m,\ell)$ that is jointly typical with $x^n$ according to $P_{UX}$, and where $u^n(m,\ell)$ is jointly typical with $y^n$ according to $P_{UY}$. This is exactly the error event in Han's scheme and has exponent $E_0(P_{U|X})$ given by \eqref{eq:EHan}.

The second error type is where all of the following happen:
\begin{enumerate}
\item[$\mE_1$:] Both $x^n$ and $y^n$ are typical respectively according to $P_X$ and $P_Y$ (this is also a necessary condition for the first error type);
\item[$\mE_2$:] The sender could find a codeword $u^n(m,\ell)$ that is jointly typical with $x^n$ according to $P_{UX}$, while $u^n(m,\ell)$ and $y^n$ have small empirical mutual information
\begin{equation}\label{eq:Ipiell}
I_{\pi^\ell}(U;Y) < R'+\delta;
\end{equation}
\item[$\mE_3$:] There exists some $k\neq \ell$ such that $u^n(m,k)$ and $y^n$ are jointly $P$-typical.
\end{enumerate}

The exponent of $\Pr[\mE_1]$ is given in \cite{shalabypapamarcou92}, but here it is more convenient to analyze $\Pr[\mE_1\cap \mE_2]$ directly. The probability for $X^n$ to be typical according to $P_X$ (when generated IID $Q_X$) has exponent $D(P_X\|Q_X)$. Conditional on $X^n$ being typical according to $P_X$, the probability for the sender to find a jointly-typical $u^n(m,\ell)$ is high due to \eqref{eq:RR'}. Given that $(u^n(m,\ell),x^n)$ are jointly typical (according to $P_{UX}$), the probability for $Y^n$ to be typical according to $P_Y$, and to have small empirical mutual information with $u^n(m,\ell)$, has exponent
 $$\min_{\tilde{P}\in \mathcal{P}^*(P_{UXY},R'+\delta)}  D\left(\tilde{P}_{Y|UX} \middle \| Q_{Y|UX}\middle| P_{UX}\right),$$
where $\set{P}^*$ is defined in \eqref{eq:defP*}.
So the exponent of $\Pr[\mE_1\cap \mE_2]$ is
\begin{IEEEeqnarray}{rCl}
\IEEEeqnarraymulticol{3}{l}{D(P_X\|Q_X) + \min_{\tilde{P}\in \mathcal{P}^*(P_{UXY},R'+\delta)}  {D\left(\tilde{P}_{Y|UX} \middle \| Q_{Y|UX}\middle| P_{UX} \right)}}\nonumber\\*
\qquad \qquad & = & \min_{\tilde{P}\in\set{P}^*(P_{UXY}, R'+\delta)} D\left(\tilde{P}_{UXY}\middle\| Q_{UXY}\right), \label{eq:minD}
\IEEEeqnarraynumspace
\end{IEEEeqnarray}
which holds by the chain rule of relative entropy and because $Q_{U|X}=P_{U|X}$. 

Given that $y^n$ is typical according to $P_Y$, the probability for a specific $U^n(m,k)$---which is generated independently of $y^n$ and IID according to $P_U$---to be jointly typical with $y^n$ according to $P_{UY}$ is approximately $2^{-nI(U;Y)}$, so $\Pr[\mE_3|\mE_1\cap \mE_2]$ has exponent
\begin{equation}\label{eq:IUY-R'}
\big(I(U;Y)-R'\big)^+.
\end{equation}
Summing  \eqref{eq:minD} and \eqref{eq:IUY-R'} and choosing $R'$ close to $I(U;X)-R$ and $\delta$ close to zero, we conclude that the optimal exponent of this second type of error is $E^*(P_{U|X},R)$ defined in \eqref{eq:EKW}.

The overall error exponent is the smaller one between the exponents of the two error types, i.e., between $E_0(P_{U|X})$ and $E^*(P_{U|X},R)$.
Optimizing this exponent over $P_{U|X}$ yields the desired result.
\end{IEEEproof}

\medskip

\begin{remark}
The above proof slightly simplifies under the same-marginal assumption~\eqref{eq:samemarginal}. Specifically, under $\mH_1$, we no longer need to consider $\mE_1$ (because it happens with high probability), whereas the events $\mE_2$ and $\mE_3$ become independent.
\end{remark}
\section{Discussions}

\subsection{Comparison with SHA}

For clearer comparison, let us again make the same-marginal assumption \eqref{eq:samemarginal}.

The main difference between SHA's expression \eqref{eq:SHAE1} and ours \eqref{eq:EKW} is: the minimization in \eqref{eq:SHAE1} is restricted to $\tilde{P}$ satisfying 
\begin{equation}\label{eq:resSHA}
I_{\tilde{P}}(U;Y) \le I_P(U;Y), 
\end{equation}
while in \eqref{eq:EKW} the restriction is 
\begin{equation}\label{eq:resnew}
I_{\tilde{P}}(U;Y) \le R'.
\end{equation}
Since SHA's scheme only works when $R'<I_P(U;Y)$, \eqref{eq:resnew} is always stronger than~\eqref{eq:resSHA}.\footnote{Our new scheme remains valid when $R'\ge I_P(U;Y)$, but one can show that in such a case the exponent is suboptimal.}

In some regimes our scheme does not improve over SHA. If $R< I_Q(U;X|Y)$, the distribution $Q_{UXY}$ is contained in both $\set{P}_\textnormal{SHA}(P_{UXY})$ and $\set{P}^*(P_{UXY},R-I(U;X|Y))$, therefore both our and SHA's receivers reduce to the na\"ive receiver. If $E_1(P_{U|X},R)>E_0(P_{U|X})$, the binning exponents in both SHA's and our schemes become inactive. 

When $R$ is in the range
\begin{IEEEeqnarray}{rCl}
\IEEEeqnarraymulticol{3}{l}{\max\left\{I_P(U;X|Y),\,I_Q(U;X|Y)\right\}} \nonumber\\*
\qquad \quad & < & R < \min\Big\{\hat{R}\colon E_1(P_{U|X},\hat{R}) \ge E_0(P_{U|X})\Big\}, 
\label{eq:good_range}\IEEEeqnarraynumspace 
\end{IEEEeqnarray}
the new scheme does improve over SHA's, in the sense that
\begin{IEEEeqnarray}{rCl}
\IEEEeqnarraymulticol{3}{l}{
\min \left\{ E_0(P_{U|X}),\,\, E^*(P_{U|X},R)\right\} } \nonumber\\*
\qquad \qquad \qquad & > &\min \left\{ E_0(P_{U|X}),\,\, E_1(P_{U|X},R)\right\}.
\label{eq:improve} \IEEEeqnarraynumspace
\end{IEEEeqnarray}
In particular, in this range, the right-hand side of \eqref{eq:improve} increases linearly with $R$ \cite{rahmanwagner12}, whereas the left-hand side increases super-linearly with $R$.

Since both SHA's and our new exponents further require maximization over $P_{U|X}$, \eqref{eq:improve} alone does not provide conclusive evidence for improvement. We next provide an example to demonstrate that the new scheme is indeed strictly better. Following Watanabe \cite{watanabe22}, we shall consider the \emph{critical rate}. This will allow us to enforce $U=X$ with probability $1$.

\subsection{An Example}
\begin{example}\label{ex:ternary}
Let $\set{X}=\{0,1,2\}$ and $\set{Y}=\{0,1\}$. Both $X$ and $Y$ are uniformly distributed under both hypotheses. Under $\mH_0$,
\begin{IEEEeqnarray}{rCl}
P_{Y|X}(1|0) & = & 0.5\\
P_{Y|X}(1|1) & = & p\\
P_{Y|X}(1|2) & = & 1-p,
\end{IEEEeqnarray}
and under $\mH_1$,
\begin{IEEEeqnarray}{rCl}
Q_{Y|X}(1|0) & = & p\\
Q_{Y|X}(1|1) & = & 0.5\\
Q_{Y|X}(1|2) & = & 1-p,
\end{IEEEeqnarray}
where $p\in(0,0.5)$. We shall compare the critical rates of Han's, SHA's, and the new scheme. The critical rate is the smallest rate that allows one to achieve the non-distributed exponent given by 
\begin{equation}\label{eq:DPQ}
D(P_{XY}\|Q_{XY}) = \frac{1}{3} \bigl( \Db(0.5\|p) + \Db(p\|0.5)\bigr),
\end{equation}
where $\Db(\cdot\|\cdot)$ denotes the relative entropy between two Bernoulli distributions of indicated parameters.
\end{example}

By \cite[Prop.~2]{watanabe22}, in this example, $E_0(P_{U|X})=D(P_{XY}\|Q_{XY})$ if, and only if, $P_{U|X}$ is such that $U=X$ with probability one. Therefore, it suffices to consider all three schemes for $U=X$. To achieve \eqref{eq:DPQ} using Han's scheme, we need a rate of
\begin{equation}\label{eq:RcrHan}
R_{\textnormal{cr-Han}} = I(X;X) = \log 3 \approx 1.59 \textnormal{ bits.}
\end{equation}

SHA's scheme reduces to the na\"ive binning scheme in this example, because 
\begin{equation}
I_P(U;X|Y) = I_Q(U;X|Y).
\end{equation}
At rate \eqref{eq:RcrHan}, it achieves 
\begin{IEEEeqnarray}{rCl}
E_\textnormal{SHA}(R_{\textnormal{cr-Han}} ) & = & R_{\textnormal{cr-Han}} - I(X;X|Y) \\
& = & I(X;Y)\\
& = & \frac{2}{3}\big(1-\Hb(p)\big) <  D(P_{XY}\|Q_{XY}), \IEEEeqnarraynumspace
\end{IEEEeqnarray}
so SHA's scheme is not useful in this example:
\begin{equation}
R_{\textnormal{cr-SHA}} > R_{\textnormal{cr-Han}}.
\end{equation}

It is easy to compute the critical rate attained by the new scheme for specific values of $p$. When $p=0.1$, we have
\begin{equation}
R_\textnormal{cr}^* \approx 1.51 \textnormal{ bits} < R_{\textnormal{cr-Han}}.
\end{equation}
Thus, the new scheme is strictly better than both Han's and SHA's schemes.

\begin{figure}[tbp]
\centering
\includegraphics[width=0.5\textwidth]{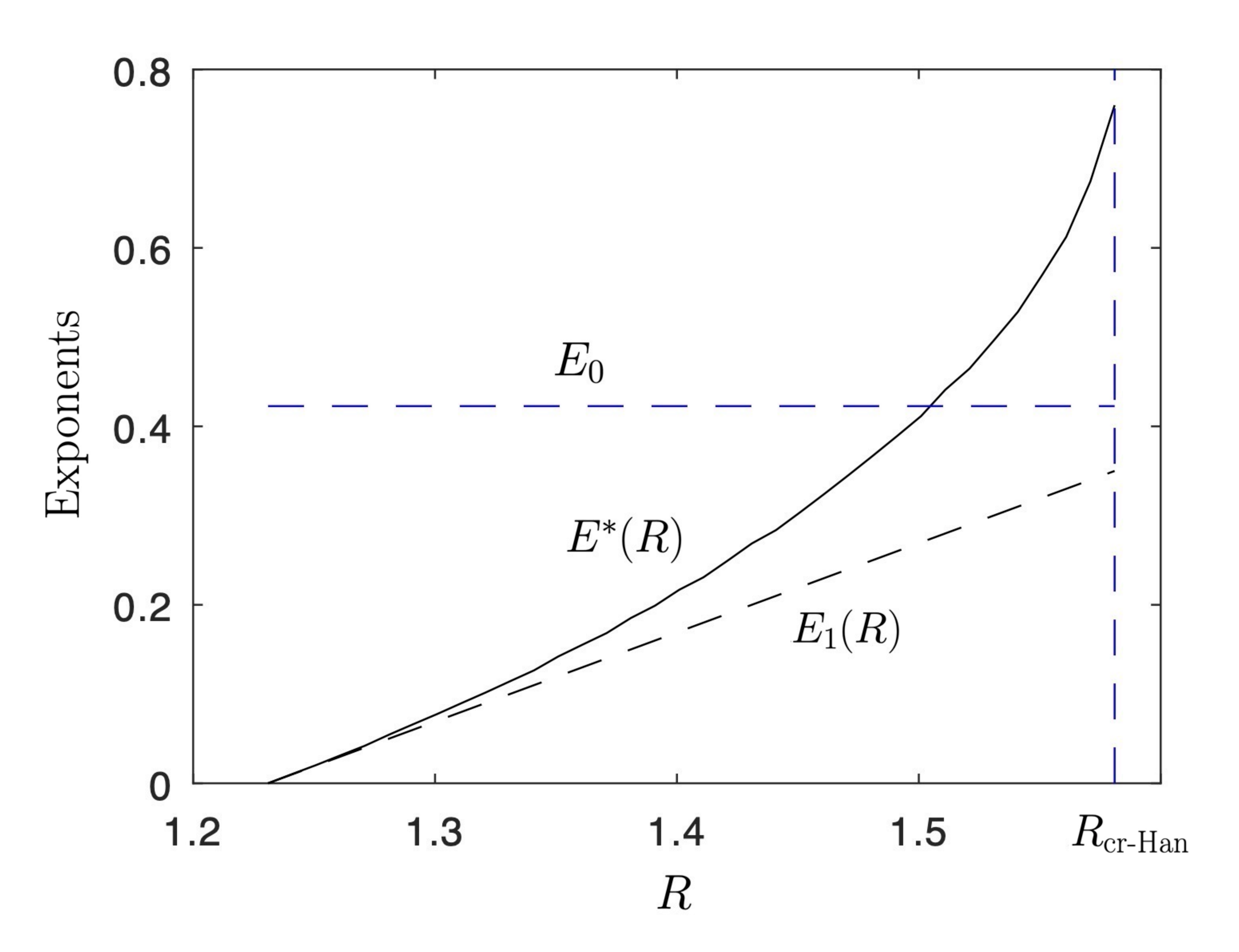}
\caption{Example~\ref{ex:ternary} with $p=0.1$ and the choice $U=X$. The exponent $E_0$ is given by \eqref{eq:EHan} and does not depend on $R$; $E_1(R)$ and $E^*(R)$ are given in \eqref{eq:SHAE1} and \eqref{eq:EKW}, respectively. Our critical rate $R_\textnormal{cr}^*$ is where $E^*(R)=E_0$. Since $E_1(R)$ lies below $E_0$ for all $R<R_\textnormal{cr-Han}$, SHA's scheme has worse performance than Han's.} 
\label{fig:ternary}
\end{figure}

In Figure~\ref{fig:ternary} we plot $E^*(R)$ and $E_1(R)$ in this example for the choice $U=X$. When $R= I(X;X|Y) \approx 1.23$ bits, $E_1(R)$ and $E^*(R)$ are both zero. As $R$ increases, as expected, $E_1(R)$ grows linearly with $R$, while $E^*(R)$ grows super-linearly. Our critical rate $R_\textnormal{cr}^*$ is where $E^*(R)$ intersects with $E_0$.

\subsection{Final Remarks}

As discussed in Section~\ref{sub:intuition}, \eqref{eq:Icond} is the optimal condition of its type, i.e., one cannot replace \eqref{eq:Icond} (with $\delta$ approaching zero) by any other condition (but keeping ``for all $k\neq \hat{\ell}$'') to achieve a larger error exponent under $\mH_1$ while still ensuring a vanishing error probability under $\mH_0$. Conditions like \eqref{eq:TV} and \eqref{eq:Du} are suboptimal because, among types that have  the same empirical mutual information, they permit some but exclude others. SHA's condition \eqref{eq:SHAcondition} is right in considering $I_{\pi^k}(U;Y)$, but suboptimal in comparing it with $I_P(U;Y)$.

We note that Watanabe's scheme in \cite{watanabe22} is not subsumed by our new scheme, because he employs separate binning by the sender, while our scheme only differs from SHA's in the receiver's decision rule. In his example, our new scheme has better performance than SHA's scheme, but does not beat his scheme. In cases where the joint distributions lend to a decomposition into two components, thus separate binning is applicable, one can combine separate binning with the new decision rule. 

\section*{Acknowledgment}

The authors wish to thank an anonymous reviewer for pointing out the connection between \cite{shimokawahanamari94} and universal decoding.


\begin{thebibliography}{10}

\bibitem{ahlswedecsiszar86}
R.~Ahlswede and I.~Csisz\'ar, ``Hypothesis testing with communication
  constraints,'' {\em IEEE Trans. Inform. Theory}, vol.~32, pp.~533--542, July
  1986.

\bibitem{han87}
T.~S. Han, ``Hypothesis testing with multiterminal data compression,'' {\em
  IEEE Trans. Inform. Theory}, vol.~33, pp.~759--772, Nov. 1987.

\bibitem{shalabypapamarcou92}
H.~M.~H. Shalaby and A.~Papamarcou, ``Multiterminal detection with zero-rate
  data compression,'' {\em IEEE Trans. Inform. Theory}, vol.~38, pp.~254--267,
  Mar. 1992.

\bibitem{shimokawahanamari94}
H.~Shimokawa, T.~S. Han, and S.~Amari, ``Error bound of hypothesis testing with
  data compression,'' in {\em Proc. IEEE Int. Symp. Inform. Theory},
  (Trondheim, Norway), June 1994.

\bibitem{rahmanwagner12}
M.~Rahman and A.~Wagner, ``On the optimality of binning for distributed
  hypothesis testing,'' {\em IEEE Trans. Inform. Theory}, vol.~58,
  pp.~6282--6303, Oct. 2012.

\bibitem{zhaolai18}
W.~Zhao and L.~Lai, ``Distributed testing with cascaded encoders,'' {\em IEEE
  Trans. Inform. Theory}, vol.~64, pp.~7339--7348, Nov. 2018.

\bibitem{weinbergerkochman19}
N.~Weinberger and Y.~Kochman, ``On the reliability function of distributed
  hypothesis testing under optimal detection,'' {\em IEEE Trans. Inform.
  Theory}, vol.~65, pp.~4940--4965, Aug. 2019.

\bibitem{salehwiggerwang19}
S.~Salehkalaibar, M.~Wigger, and L.~Wang, ``Hypothesis testing over the two-hop
  relay network,'' {\em IEEE Trans. Inform. Theory}, vol.~65, pp.~4411--4433,
  July 2019.

\bibitem{salehwigger20}
S.~Salehkalaibar and M.~Wigger, ``Distributed hypothesis testing based on
  unequal-error protection codes,'' {\em IEEE Trans. Inform. Theory}, vol.~66,
  pp.~4150--4182, July 2020.

\bibitem{sreekumargunduz20}
S.~Sreekumar and D.~Gündüz, ``Distributed hypothesis testing over discrete
  memoryless channels,'' {\em IEEE Trans. Inform. Theory}, vol.~66,
  pp.~2044--2066, Apr. 2020.

\bibitem{watanabe22}
S.~Watanabe, ``On sub-optimality of random binning for distributed hypothesis
  testing,'' in {\em Proc. IEEE Int. Symp. Inform. Theory}, (Espoo, Finland),
  June 2022.

\bibitem{wynerziv76}
A.~D. Wyner and J.~Ziv, ``The rate-distortion function for source coding with
  side information at the decoder,'' {\em IEEE Trans. Inform. Theory}, vol.~22,
  pp.~1--10, Jan. 1976.

\bibitem{coverthomas06}
T.~M. Cover and J.~A. Thomas, {\em Elements of Information Theory}.
\newblock New York: John Wiley \& Sons, second~ed., 2006.

\bibitem{elgamalkim11}
A.~El~Gamal and Y.-H. Kim, {\em Network Information Theory}.
\newblock Cambridge University Press, 2011.

\bibitem{csiszarkorner11}
I.~Csisz\'{a}r and J.~K\"{o}rner, {\em Information Theory: Coding Theorems for
  Discrete Memoryless Systems}.
\newblock Cambridge University Press, second~ed., 2011.

\end{thebibliography}
\end{document}